\documentclass[fleqn,usenatbib]{mnras}
\usepackage{newtxtext,newtxmath}
\usepackage[T1]{fontenc}
\usepackage{ae,aecompl}
\usepackage{graphicx}	
\usepackage{amsmath}	
\usepackage{amssymb}	
\usepackage{hyperref}

\title[Local density distribution of the Galaxy]{A method to calculate the local density distribution of the Galaxy from the Tycho-Gaia Astrometric Solution data}

\author[Rain Kipper, Elmo Tempel and Peeter Tenjes]{Rain Kipper$^{1}$\thanks{E-mail:
rain@to.ee}, Elmo Tempel$^{1,2}$ and Peeter Tenjes$^{1}$\\
$^{1}$Tartu Observatory, Observatooriumi 1, 61602 T\~oravere, Estonia\\
$^{2}$Leibniz-Institut f\"ur Astrophysik Potsdam (AIP), An der Sternwarte 16, 14482 Potsdam, Germany
}

\begin{document}

\date{Accepted 2017 September 19. Received 2017 September 19; in original form 2017 June 29}
\pagerange{\pageref{firstpage}--\pageref{lastpage}} \pubyear{2002}
\maketitle
\label{firstpage}
\begin{abstract}	
New and more reliable distances and proper motions of a large number of stars in the Tycho-Gaia Astrometric Solution (TGAS) catalogue allow to calculate the local matter density distribution more precisely than earlier. 

We devised a method to calculate the stationary gravitational potential distribution perpendicular to the Galactic plane by comparing the vertical probability density distribution of a sample of observed stars with the theoretical probability density distribution computed from their vertical coordinates and velocities. We applied the model to idealised test stars and to the real observational samples. Tests with two mock datasets proved that the method is viable and provides reasonable results.

Applying the method to TGAS data we derived that the total matter density in the Solar neighbourhood is $0.09\pm 0.02~\text{M}_\odot\text{pc}^{-3}$ being consistent with the results from literature. The matter surface density within $|z|\le 0.75~\text{kpc}$ is $42\pm 4~\text{M}_\odot\text{pc}^{-2}$. This is slightly less than the results derived by other authors but within errors is consistent with previous estimates. Our results show no firm evidence for significant amount of dark matter in the Solar neighbourhood. However, we caution that our calculations at $|z| \leq 0.75$~kpc rel on an extrapolation from the velocity distribution function calculated at $|z| \leq 25$~pc. This extrapolation can be very sensitive to our assumption that the stellar motions are perfectly decoupled in R and z, and to our assumption of equilibrium. Indeed, we find that $\rho (z)$ within $|z|\le 0.75$~kpc is asymmetric with respect to the Galactic plane at distances $|z| = 0.1-0.4$~kpc indicating that the density distribution may be influenced by density perturbations.

\end{abstract}

\begin{keywords}
Galaxy: fundamental parameters -- Galaxy: kinematics and dynamics -- Solar neighbourhood -- stars: kinematics and dynamics -- methods: data analysis
\end{keywords}
	
\section{Introduction} \label{sec:introduction}
    
Rather accurate distances and proper motions of a large number of stars in the Tycho-Gaia Astrometric Solution (TGAS) \citep{Gaia1, Gaia2, Lindegren:2016} catalogue allow to calculate the Milky Way local matter density in the Solar neighbourhood more precisely than it was possible earlier. Moreover, it is possible to study the variations of spatial matter densities near to the Galactic plane and thus to study possible density distribution asymmetries with respect to the Milky Way plane. 
    
Local matter density is an important parameter allowing to constrain possible mass distribution models of the Galaxy \citep[see e.g.][]{Dehnen:1998, Olling:2001} and to derive constraints to the local dark matter density \citep{Read:2014}. When knowing the dark matter density or having certain constraints to its value it is possible to estimate limits for the dark matter particle annihilation cross section \citep[e.g.][]{OHare:2016, Stref:2017} and thus to derive constraints to the nature of dark matter particles.
    
Local mass density determination has been in focus for astronomers at least since a classical paper by \citet{Oort:1932}. Initially the local mass density was handled as a parameter allowing to limit Milky Way overall mass distribution models \citep{Kuzmin:1952, Kuzmin:1955, Oort:1960}. In 80's attention moved to the determination of the amount of the local dark matter \citep[see review by][]{Einasto:2005}. An important step in determining the matter density in the Solar neighbourhood was done by \citet{Kuijken:1989a, Kuijken:1989b, Kuijken:1989c}. By assuming (similar to \citealt{Oort:1932}) that near to the galactic plane the gravitational potential is a function of $z$ coordinate only\footnote{$(R, \theta, z)$ are galactocentric cylindrical coordinates, where $z$ is perpendicular to a galactic plane and $z=0$ corresponds to the plane used in definition of galactic coordinates ($l$, $b$).}, $\Phi (R,\theta , z) \simeq \Phi (z)$, the phase density of stars is only a function of $z$ energy, $f(E_z)$, and applying the method to a sample of about 500 K-type dwarf stars they calculated the surface density of matter within a certain distance from the Galactic plane. At the same time an important paper by \citet{Statler:1989} was published where a model was developed without using these assumptions for the gravitational potential and for the phase density distributions. Assuming that the gravitational potential of the Milky Way disc can be approximated with the St\"ackel potential, \citet{Statler:1989} analysed kinematic properties of stars at different distances from the Galactic plane and demonstrated that classical simplifying assumptions $\Phi = \Phi (z)$ and $f = f(E_z)$  can be used only up to the distances $|z| \simeq 1~\text{kpc}$. In this case the third integral of motion is $I_3 = E_z$. Taking into account this limit the method developed by \citet{Kuijken:1989a} were used for different samples about a similar number of stars by e.g. \citet{Kuijken:1991} and \citet{Flynn:1994}. Calculated matter surface densities were $\Sigma (|z| \le 1.1~\text{kpc}) = 71$ and $52~\mathrm{M}_\odot\mathrm{pc}^{-2}$, respectively, and a general conclusion was made that the amount of dark matter in the Solar neighbourhood is rather small. 

The limit $|z| \simeq 1~\text{kpc}$ should be handled with caution since \citet{Garbari:2011} derived that even at $|z|\simeq 0.5~\text{kpc}$ the assumption about decoupled $z$ motions may not be a good approximation (see Sect.~\ref{sec:discussion} for more discussion).
    
When parallaxes and proper motions for a large number of nearby stars measured by the \textit{Hipparcos} satellite \citep{ESA:1997} became available, these high-quality data were used to determine local matter densities in the Solar neighbourhood. By using \textit{Hipparcos} data it was possible to eliminate several errors that affected the precision of earlier density estimates. However, as \textit{Hipparcos} measured parallaxes with sufficient accuracy only up to about $\sim\!130~\text{pc}$ from the Sun, these data allowed to determine the local spatial density only and not the surface densities up to larger distances from the Galactic plane. Following the release of \textit{Hipparcos} data local matter density values were determined e.g. by \citet{Creze:1998}, \citet{Holmberg:2000, Holmberg:2004}, \citet{Siebert:2003}, and \citet{Bienayme:2006}.

When using \textit{Hipparcos} data an additional problem arises. For most of the stars measured by \textit{Hipparcos} radial velocity data are not available or these are not sufficient quality. For \textit{Hipparcos} stars only the proper motions are homogeneous and precise enough. Due to this it is needed to identify the measured proper motion (tangential) velocities with the $z$ component velocities. This approximation is valid only at low Galactic latitudes and limits $|b| \le \mathrm{10\degr - 12\degr}$ were used by \citet{Creze:1998} and \citet{Holmberg:2000} in their analysis. This decreases the number of used stars in their data samples to about 500--1300.

In recent years several datasets were combined to have larger samples of stars with sufficient observational data. \citet{Garbari:2011, Garbari:2012} added to the sample of K-type dwarfs used by \citet{Kuijken:1989b} their own velocities and also stellar distances measured by \textit{Hipparcos} and SEGUE\footnote{Sloan Extension for Galactic Understanding and Exploration \citep{Yanny:2009}.}. Additional spectral, proper motion and/or radial velocity data were also used by \citet{Siebert:2003}, \citet{Bienayme:2006}, \citet{Bovy:2013}, \citet{Zhang:2013}, and \citet{Bienayme:2014}.  Additional data allowed to reduce statistical uncertainties in calculated density estimates. 

As we mentioned earlier, to calculate the local matter density, it is usually assumed that gravitational potential is a function of $z$ coordinate only, $\Phi (z)$, and phase density of stars is a function of $z$ energy only, $f(E_z)$. Thereafter, certain analytical forms for $f(E_z)$ and $\Phi (z)$ are selected. Phase density is assumed to consist of several components with known velocity dispersions and representing different galactic populations. For the gravitational potential, also a suitable analytical form is selected. It is possible also to start from selection of an analytical form for the mass density distribution, e.g., by assuming the Galaxy to consist of a superposition of several isothermal components with known velocity dispersions. In this case the phase density can be determined empirically by fitting the observed velocity distribution of stars with some analytical function. Thereafter, by using of the known equation relating the phase density and spatial density, predicted density distribution $\rho (z)$ of a used sample of stars can be calculated. By fitting the predicted and observed density distributions it is possible to determine free parameters in selected gravitational potential. As the gravitational potential and mass density are related via the Poisson equation it is possible to calculate the total mass density distribution.

In the present study, we develop a method to calculate the local mass density distribution in the Solar neighbourhood by using coordinates and velocities of a large number of individual stars from the TGAS data. Our model is somewhat different from the previously described approaches. We start from the conventional assumption that the gravitational potential is in form of $\Phi (z)$ and that stellar orbits are vertical oscillations. Thus, the method is applicable near to the Galactic plane only. But in our model, we sum the vertical orbits, calculate the probability of a star to be located between $z$ and $z+\Delta z$ and compare these probabilities with their observed vertical distribution. As a consequence, the free parameters for the gravitational potential can be determined. 

The method developed in this paper is presented in Sect.~\ref{sec:model}. In Sect.~\ref{sec:mock} we generate several mock datasets and test our model using these mock datasets. After concluding that the developed model allows to determine true parameters of the mock datasets with sufficient precision the model is applied to the real data in Sect.~\ref{sec:application_of_solar_region}. In Sect.~\ref{sub:data} selection of the data from TGAS and formation of a volume limited data sample is described. Modelling process and results of model calculations are presented in Sections~\ref{sub:fitting} and \ref{sub:results}. In Sect.~\ref{sec:discussion} we discuss our results.

	
	\section{Model} \label{sec:model}
	Our aim in the present paper is to study Milky Way matter density distribution perpendicular to the Galactic plane in the Solar neighbourhood. We assume the gravitational potential to be stationary. As we limit ourselves also to the regions nearby to the Galactic plane and in the Solar neighbourhood ($\partial/\partial R = \partial/\partial\theta =0$) we can assume the gravitational potential to be in form of $\Phi = \Phi(z)$. In this case, the $z$-directional energy is the third integral of motion $I_3 = E_z = E$ \citep[see e.g.][]{Statler:1989}. Phases of stars in their orbits are assumed to be random. 
	
	\subsection{General outline of the method}
	Instead of working with stellar number densities, we construct our model based on probability densities. Let $p(z)\text{d}z$ be the probability to find a star between $z$ and $z+\text{d}z$ calculated from the model. The same quantity, but calculated directly from the observed $z$ coordinate distribution for a sample of stars is denoted as $p_{\text{obs}}(z)\text{d}z$. In case of a self-consistent model these probabilities should be equal for all $z$, i.e.
	\begin{equation}
		p(z)\mathrm{d}z = p_{\mathrm{obs}} (z)\mathrm{d}z \label{eq:fittimise_ideaal}. 
	\end{equation}
	
	We assume in our model that a sample of test stars move in an overall gravitational potential of a galaxy. Within our approximations we can assume that stellar orbits are simply vertical oscillations. The calculated probability density $p(z)$ for the sample of test stars consists of probability densities of individual stellar orbital probabilities and can be written as a weighted sum of individual orbital probability densities $p_i(z)$:
	\begin{equation}
		p(z) = \frac{\sum w_i p_i(z)}{\sum w_i}. \label{eq:orbiitide_summa}
	\end{equation}
	The summation is over different orbits ($i$) corresponding to individual stars. At present stage, there is no reason to prefer one orbit to other and thus all orbits/stars have equal weights ($w_i = 1$).
	
	To calculate the probability density for an orbit ($i$), we may assume that the probability to find a star between $z - \text{d}z/2$ and $z + \text{d}z/2$ is proportional to the time it spends there
    \begin{equation}
    p_i (z)\mathrm{d}z = \frac{\mathrm{d}t}{T_i} = \frac{1}{T_i} \frac{\mathrm{d}z}{v_i(z)},   
	\label{eq:pi_valem}
    \end{equation}
	where $T_i$ is the semi-period of the orbit $i$ (i.e. normalising constant) and $v_i (z)$
    is the velocity of a star ($i$) in $z$ direction. Velocities in the vertical direction $v(z)$ can be found from the conservation of the $E_z$ energy integral:	
	\begin{equation}
		v(z) = \sqrt{2\Phi(z_0) + v_{0}^2 - 2\Phi (z)}. \label{eq:v_from_pot}
	\end{equation}	
	The quantities $v_0$ and $z_0$ are the $z$ component velocity and position of a star at a particular moment. For all stars, these values can be taken to be equal to the values given in an observed database of stars.  Illustration of the probability densities for some orbits and their sum is given in Fig.~\ref{fig:orbit_illustration}. 
	
	Thus, if we have a database of stars with known coordinates and velocities at present epoch we may calculate their observed coordinate distribution $p_{\text{obs}}(z)$ and on the other hand by choosing a sufficiently flexible analytical form for gravitational potential $\Phi(z)$ we may calculate $p(z)\text{d}z$ from Eq.~(\ref{eq:orbiitide_summa}). By fitting $p(z)$ to $p_{\text{obs}} (z)$ we can calculate gravitational potential parameters. 
	
	The method is similar to the one-dimensional Schwarzschild galaxy modelling method \citep{Schwarzschild:1979}. Observed phase-space coordinates of our sample stars $(z_0, v_0)$ correspond to initial conditions in the orbit library calculations of the Schwarzschild's method. As a difference from the Schwarzschild's method, we do not need to fit weights of different orbits.

	\subsection{Requirements for data}\label{sec:model_assumptions}
    	
As we compare $p(z)$ and $p_{\text{obs}} (z)$ it is required that the energy distributions of stars used to calculate $p_\mathrm{obs} (z)$ and $p (z)$ are the same. If the samples of stars are the same this requirement is automatically fulfilled. But if the samples are different, e.g. if we use only a subsample of stars to calculate $p(z)$ compared to the larger sample to calculate $p_\mathrm{obs}$ (see Sect.~\ref{sec:tracer}), then this requirement is not fulfilled by itself. 
	
\begin{figure}
    \includegraphics{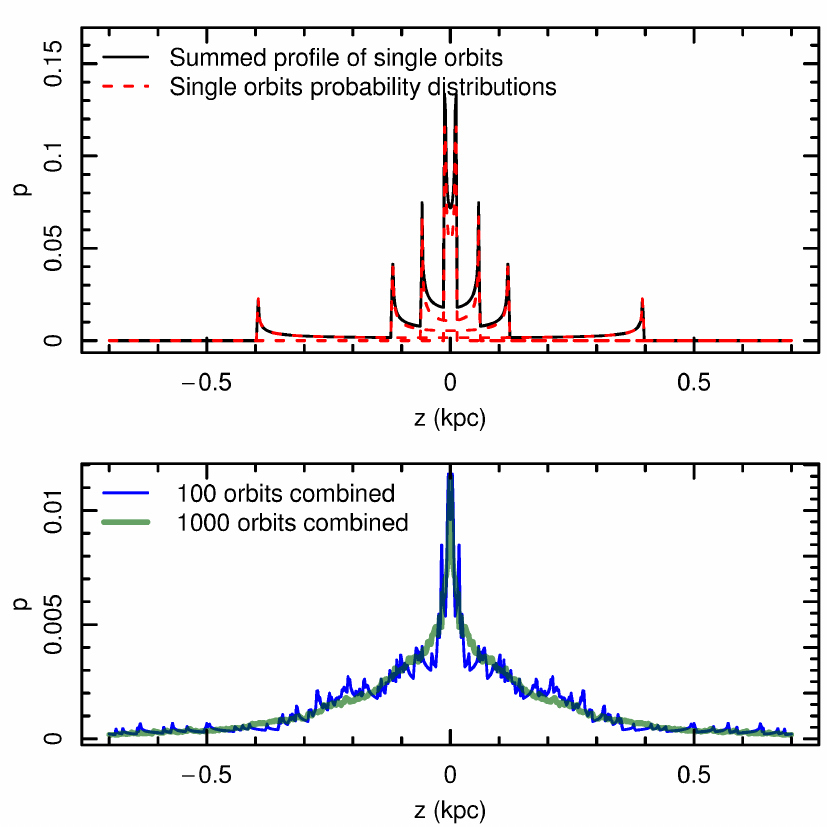}
    \caption{Probability density to find a star  at $z$ for a sample consisting of four stars (upper panel). The dashed red lines indicate probability distributions $p_i$ for individual orbits, continuous black line gives their sum $p(z)$. The lower panel shows that the model line $p(z)$ becomes a smoother function when the number of orbits increases. }\label{fig:orbit_illustration} 
    \end{figure}

	Individual stellar probability densities ($p_i$) have a specific spiky shape (see Fig.~\ref{fig:orbit_illustration}). For this reason, we decided to bin our probability distributions to smooth out spikes of individual stars.
		
	\subsection{Tracer region}\label{sec:tracer}
	In some cases, velocities (proper motions and/or radial velocities) are available for a different (usually smaller) sample of stars when compared to stellar distances. To calculate $p_\mathrm{obs}(z)$ only stellar distances are needed, to calculate $p(z)$ both the distances and velocities are needed. Thus, the stars used to calculate $p(z)$ and the stars used to construct $p_\mathrm{obs}(z)$ need not to be the same. But as we mentioned already in Sect.~\ref{sec:model_assumptions} they need to have the same energy distribution. In our model, we add a possibility to take this selection effect into account when calculating $p(z)$ distribution.    
	
	Let $f_\mathrm{tr}(z)$ be a selection function, being a probability for a star to be identified as a tracer star and to be used to calculate the model $p(z)$. Usually, the $f_\mathrm{tr}$ can be considered as known function. In most cases, $f_\mathrm{tr}$ is less or equal than one, i.e. the tracer stars form a subset of comparison stars. 
	
	For each orbit, the probability that this star is part of a tracer population can be calculated from $f_\mathrm{tr}$ and $p(z)$: 
	\begin{equation}
		P_{\mathrm{tr},i} = \frac{ \int_{-\infty}^\infty p_i(z)f_\text{tr}(z) \text{d}z}{ \int_{-\infty}^\infty  p_i(z)\text{d}z}.
	\end{equation}
	Hence, for each orbit, tracer stars are under-represented by a factor of $P_{\text{tr},i}$. If we want to construct $p(z)$ from tracer stars, we need to compensate this under-representation by weighting up these orbits in Eq.~(\ref{eq:orbiitide_summa}) by the same factor
	\begin{equation}
		w_i = \frac{1}{P_{\mathrm{tr},i}} . \label{eq:w}
	\end{equation}
If the selection function is a function of $z$ only, the compensation in form of Eq.~(\ref{eq:w}) conserves the energy distribution assuming that the number of tracer stars is sufficient. A test about the quality of the energy distribution conservation is seen in Sect. \ref{sec:mock_tests} and Fig.~\ref{fig:energy_distr}.
	
	To apply the tracer star concept $f_\mathrm{tr}$ we formulated some requirements for the construction of the tracer population. First, to ensure that all orbits pass tracer region and to include stars with all possible energies, the tracer stars are suggested to be located near to the minimum of the gravitational potential. In case, where the stars are located away from the minimum, the potential can be determined only above the tracer star region, i.e. above some limiting height. Second, the tracer stars must have known velocities and positions. Third, the number of tracer stars should be as large as possible to suppress spiky nature of $p(z)$. 
	
	The simplest case for selecting tracer stars is to choose only stars that are below some limit ($z_\mathrm{tr}$), e.g.
	\begin{equation}
		f_\mathrm{tr} = 
		\begin{cases} 
		      0 & |z| > z_\mathrm{tr} \\          
			  1 & |z|\leq  z_\mathrm{tr}. \\
		\end{cases}\label{eq:simple_tracer}
	\end{equation}
	A more general case is 
	\begin{equation}
		f_\mathrm{tr}(z) \sim \frac{p_\mathrm{tr}(z)}{p_\mathrm{obs}(z)},
	\end{equation}
	where $p_\mathrm{tr}(z)$ is the distribution created from tracer stars. Using these formulae, we assume that there is no bias created by selecting stars for $p_\mathrm{tr}$ (e.g. line-of-sight velocity measurements only for brighter/specific population of stars).

	\subsection{Fitting process}
    \label{sec:fitting}
    
	When fitting calculated probability distribution of stars $p(z)$ to $p_\mathrm{obs}(z)$ of the comparison stars, we bin these distributions with a finite step $\Delta z$. Binning helps to smooth out the spiky nature of $p(z)$. This in turn brings us to the need to linearize the gravitational potential around the centre of a coordinate bin $z_j$. In case of finite bin widths the velocities are not constant in different bin parts (especially in regions, where stellar velocities are small but accelerations are significant), and therefore we must include an acceleration term in Eqs.~\eqref{eq:pi_valem} and \eqref{eq:v_from_pot}. Thus, the expression to calculate the velocities will have the form:
	\begin{equation}
		v(z) = \sqrt{2\Phi_0+v_0^2-2\left[\Phi(z_j) +  \left.  \frac{\text{d}\Phi}{\text{d}z} \right\rvert_{z_j} (z-z_j)\right]}
	\end{equation}
	and $\mathrm{d}t$ in Eq.~(\ref{eq:pi_valem}) will become
    \[ 
    \Delta t = \int_{-\Delta z/2}^{\Delta z/2} \frac{\text{d}z}{v} =
    \]
    \begin{equation}
    = \left( \left.\frac{\text{d}\Phi}{\text{d} z} \right\rvert_{z_j} \right)^{-1}
    \left.\sqrt{2(\Phi(z_0) - \Phi(z_j)) + v_0^2- \left.2\frac{\text{d}\Phi}{\text{d}z} \right\rvert_{z_j}  z}\,\,\right\rvert_{-\Delta z/2}^{\Delta z/2}. \\
     \end{equation}
	
	One specific feature of the model is that the model line $p(z)$ is not smooth (see Fig.~\ref{fig:orbit_illustration}). Smoothness is desired when fitting the relation (\ref{eq:fittimise_ideaal}). It can be achieved either by increasing $\Delta z$, by increasing the tracer sample size or by calculating the $p$ and $p_\mathrm{obs}$ only up to smaller height values. 
    
	In the first case, by increasing of $\Delta z$ individual spikes are suppressed but the spatial resolution of the calculated gravitational potential will be worse. The second case is clearly most preferable but in most cases no additional data are available. The third case was to change the fitting limits i.e. to remove outer regions -- as there are less stars in outer regions, the spikiness in their probability distributions has more severe influence there. In doing so a drawback is that in this case we remove high energy stars which need largest negative acceleration from the disc and just these stars are needed to constrain the overall surface density within a given region. 	

If the spiky model line persists, the fitting of $p$ to $p_\mathrm{obs}$ is influenced by having many local optima. Another way to mitigate the spikiness is to include an error term for model distribution (or equivalently increase the errors of $p_\mathrm{obs}$ artificially). One should bear in mind, that in this case the resulting errors of the fit are not true ones, but overestimated. 
	
    To fit $p_\mathrm{obs} (z)$ and $p(z)$ we first calculated the values of $p_\mathrm{obs} (z_j)$ and their uncertainties $\Delta\,[p_\mathrm{obs}(z_j)]$ at bins $z_j-\Delta z/2 < z < z_j + \Delta z/2$ and the values of $p(z_j)$ from Eq.~\eqref{eq:orbiitide_summa} at the same bins.  Fitting of these two quantities is done by minimising 
    \begin{equation}
    \chi^2 = \sum_{z_j}\frac{[p(z_j)-p_\text{obs}(z_j)]^2}{[\Delta p_\text{obs}(z_j)]^2}.
    \label{eq:chisq}
    \end{equation}


	\section{Testing the method with mock data}\label{sec:mock}
	It is common to verify methods in order to be applicable, therefore we constructed mock datasets for the verification. The aim is to test the method in different conditions, hence we created two mock sets for two general case. 
	
	\subsection{Creating the mock data}\label{sec:mock_simul}
	For both mock datasets, we assumed that gravitational potential and associated mass density are the same functions of $z$ coordinate. We chose the density distribution 
    \begin{equation}
    \rho (z) = \frac{\Sigma_0}{2h} \text{sech}^2 \,\left( \frac{z-z_0}{h}\right)
    \label{eq:sech}
    \end{equation}
    and corresponding potential
    \begin{equation}
		\Phi (z) =  2\pi G \Sigma_0 h\log\left[\cosh\left(\frac{z-z_0}{h}\right)\right].
    	\label{eq:sech_pot}
    \end{equation}
	The free parameters for the density and the gravitational potential were scale height $h = 300~\text{pc}$, surface density $\Sigma_0 = 80~\mathrm{M}_\odot\mathrm{pc}^{-2}$ and centre coordinate $z_0 = 0$~pc. This is the potential that we try to recover with our proposed method. 
	
	The distribution of test-particles (stars) were chosen randomly according to $\mathrm{sech}^2$ (Mock~1) or Gaussian (Mock~2) laws. Both mocks contained 100\,000 stars. For Mock~1 the free parameter value was taken $h = 300~\text{pc}$, and for Mock~2 Gaussian scale height was taken $h = 600~\text{pc}$. For both mock datasets, we also needed to assign velocities for each test-particle. This was done in the following way. First, from the test-particle density distribution $\tilde{\rho}$ ($\mathrm{sech}^2$ or Gaussian) and the total gravitational potential (see Eq.~(\ref{eq:sech_pot})), velocity dispersions for a given test-particle population were calculated by using one-dimensional Jeans equation	
    \begin{equation}
    \sigma^2(z) = \frac{1}{\tilde{\rho}(z)}\int\limits_{z}^{\infty}\tilde{\rho}(z^{\ast})\Phi_z (z^{\ast})\text{d}z^{\ast}.\label{eq:jeans}
    \end{equation}
	Here $\Phi_z$ denotes the derivative of the gravitational potential. Thereafter, random velocities were selected according to a normal distribution with zero mean and dispersions calculated from Eq.~(\ref{eq:jeans}). 
	
	In reality, the velocity distribution does not follow normal distribution, it depends on the gravitational potential and normal distribution should be treated as a first approximation only. To have individual stellar coordinates and velocities consistent with gravitational potential and to have mixed phases of stars in their orbits we simulated the positions and velocities of our test-stars for 10~Gyr and thereafter used this snapshot. As expected, the velocity and position distribution did not remain exactly the same. Changes in velocity distribution were up to 9 per cent and in density distribution up to 6 per cent.
	
\begin{figure}
	\includegraphics{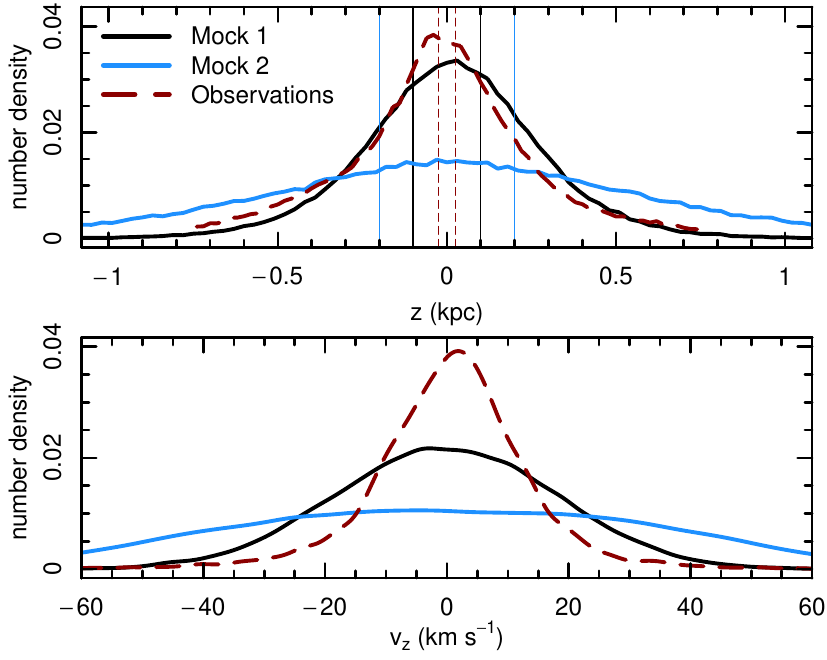}
	\caption{The upper panel shows the distribution of $z$ coordinates of 100\,000 mock test stars. Mock~1 coordinates were generated according to the $\text{sech}^2$ distribution, Mock~2 coordinates according to the Gaussian distribution (see text for details). Vertical lines mark tracer regions used for these mock data. In the lower panel velocity distributions of the test stars within the tracer regions for Mock~1B and~2B are given (see Table~\ref{tab:mock_nimed}). Red lines correspond to the sample of observed stars (see Sect.~\ref{sub:data}) that are shown for comparison. The distributions are given in arbitrary units.}
	\label{fig:mock_data}
	\end{figure}

	The final distributions of coordinates for Mock~1 and Mock~2 are given in the upper panel of Fig.~\ref{fig:mock_data}. It is seen that these two mock datasets have rather different coordinate distributions, although the underlying gravitational potential is the same. For a comparison, also the distribution of coordinates of real observational data (see Sect.~\ref{sub:data}) is given. 
	
	When using of real data, velocity information is available only for a smaller tracer region (limited by $z_\text{tr}$ value) sample. Velocity distributions for Mock~1 and Mock~2 datasets within their tracer regions are given in the lower panel of Fig.~\ref{fig:mock_data}. For a comparison, the distribution of velocities of real observational data within tracer region is given.
		
	\subsection{Modelling the mock data}
	Modelling of the mock data was done in a similar way for both datasets. We fitted $p(z)$ to $p_\mathrm{obs}(z)$ using Eq.~\eqref{eq:chisq} and defined log-likelihood as following
	\begin{equation}
		L_{\chi^2} =  - \sum_{z_j} \frac{\left[p(z_j)-p_\mathrm{obs}(z_j)\right]^2}{2\Delta p_\mathrm{obs}(z_j)^2} \label{eq:ll_chisq}
	\end{equation}
	We chose the tracers in the form of Eq.~\eqref{eq:simple_tracer}, with the parameters for each subsample given in Table~\ref{tab:mock_nimed}. The $\Delta z$ were $8$ and $20$~pc for Mock~1 and~2, respectively. 
	
	The comparison distribution $p_\mathrm{obs} (z)$ was constructed as a histogram, with the uncertainties in each bin taken according to the Poisson distribution. When noted so, these uncertainties were artificially increased. 
	
	The fitting process was done by using Bayesian interface with MCMC Metropolis-Hastings sampling. Free parameters of the gravitational potential were the scale height $h$, the total surface density $\Sigma_0$ and the central location of the gravitational potential distribution $z_0$. Since the central location is very well fitted with thin posterior distribution, we discard it from further analysis. 
The prior values for each fitting were the same: uniform priors with the range of $h\in\{ 100 \dots 500 \}$~pc and $\Sigma_0\in\{ 20 \dots 130 \}~\mathrm{M}_\odot\mathrm{pc}^{-2}$. The different modelling parameters for each mock set can be found from Table~\ref{tab:mock_nimed}.

	\begin{table}
	\caption{General parameters of mock datasets. Total number of stars (datapoints) in each mock sample is 100\,000.}
	\label{tab:mock_nimed}
	\center
	\begin{tabular}{lcccc}
	\hline\hline
	 Name       & Tracer        & Added          & Added  & Number of      \\
	            & region        & $p_\text{obs}$ & $v_z$  & tracer     \\
	            & $z_\text{tr}$ & errors         & errors & datapoints \\
	\hline
	Mock 1A   & 0.8 kpc & 0\%  & 0 & 99\,024\\
	Mock 1B   & 0.1 kpc & 0\%  & 0 & 32\,146\\
	Mock 1C   & 0.1 kpc & 25\% & 0 & 32\,146\\
	Mock 2B   & 0.2 kpc & 0\%  & 0 & 28\,032\\
	Mock 2C   & 0.2 kpc & 25\% & 0 & 28\,032\\ 
	Mock 2D   & 0.2 kpc & 25\% & 2~$\text{km\, s}^{-1}$ & 28\,032\\
	\hline\hline
	\end{tabular}
	\end{table}

	\subsection{Tests}\label{sec:mock_tests}
	
			We tested how well the modelling restored the true gravitational potential parameter values of the initial disc model. 
	
\begin{figure}
			\includegraphics{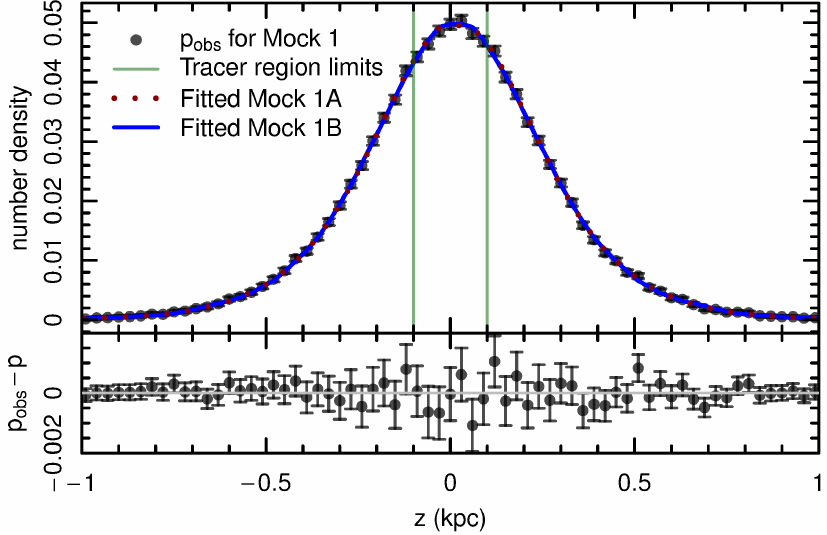}
			\caption{The fitted $p(z)$ to $p_\mathrm{obs}(z)$ for Mock~1A and Mock~1B. The top panel shows the test-particle distribution for Mock~1A and 1B, the bottom panel shows residuals for Mock~1B. }\label{fig:mock1_res}
		\end{figure}

\begin{figure}
			\includegraphics{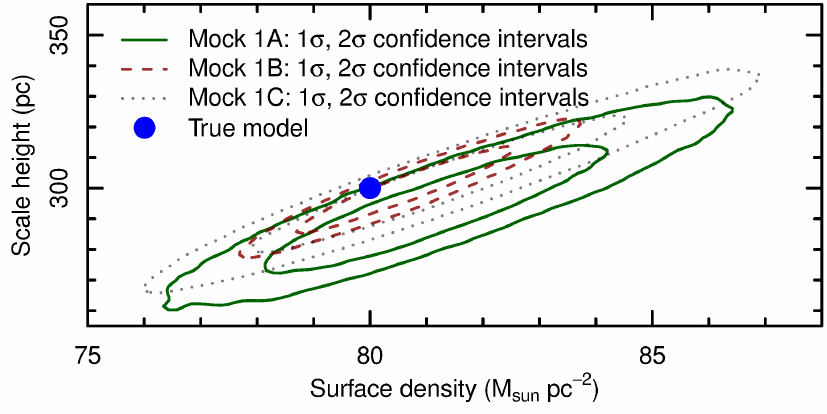}
			\caption{The posterior distributions of the fitted parameters of different Mock~1 datasets.  }\label{fig:mock1_post}
		\end{figure}

		First, we modelled Mock 1 data for the simplest case, where all the dataset stars were included to the tracer region, i.e. $z_\text{tr} = 0.8~\text{kpc}$ (Mock~1A). The precision, how well the model $p(z)$ follows $p_\mathrm{obs}(z)$ can be seen in Figs.~\ref{fig:mock1_res} and \ref{fig:mock1_post}. We conclude, that the true values were restored well and there are no systematic biases.

\begin{figure}
			\includegraphics{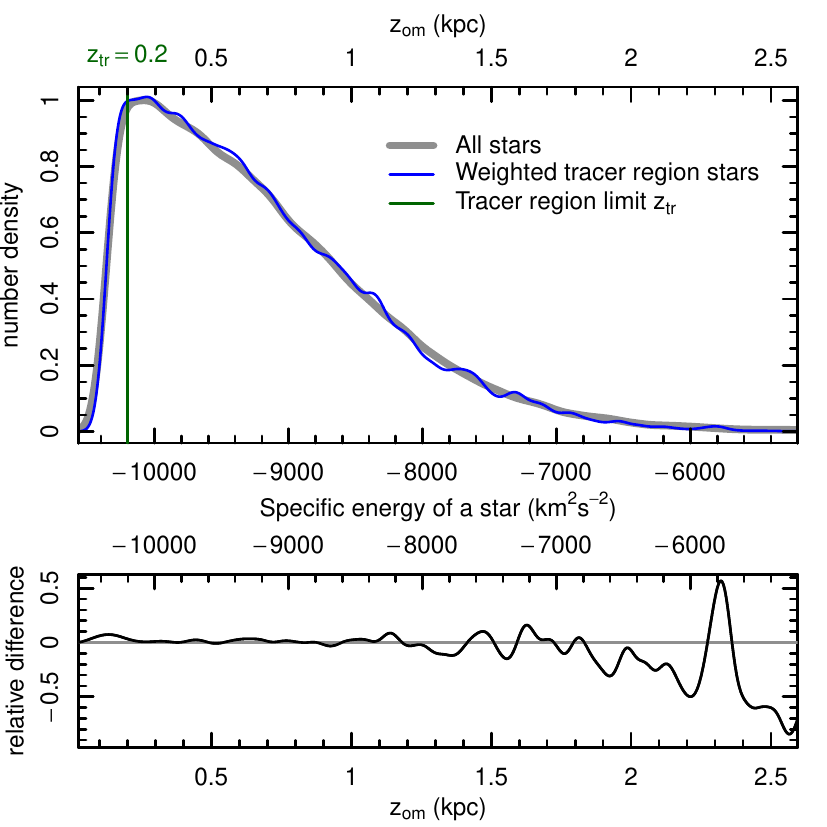}
			\caption{Energy $E_z$ (and corresponding oscillation amplitude $z_\text{om}$) distribution of stars for Mock~2 data (upper panel). The thick grey line gives the energy distribution of all stars, the blue line only for the tracer region stars weighted according to Eq.~\eqref{eq:w}. In lower panel the relative difference between these two distributions is given. The points with oscillation amplitude higher than tracer limit ($z_\text{om} > z_\text{tr}$), have $P_\text{tr} < 1$ and can produce enhanced spikiness in the modelled $p(z)$ distribution.   }\label{fig:energy_distr}
		\end{figure}

		Second, smaller tracer regions $z_\text{tr} = 0.1$ or $0.2\,\text{kpc}$ were fixed (Mock~1B and others). A general assumption for this modelling (see Sect.~\ref{sec:model_assumptions}) is that the energy distribution had to be the same for weighted tracer population and for the overall energy distribution. Calculated energy distribution for these two particle samples is given in Fig.~\ref{fig:energy_distr}. One can see some deviations, mostly in high energy region, which corresponds to higher regions above the plane. The region, where the fluctuations are high, should be used with caution. As the energy distributions match, we can check how does the modelling performs in a case of a smaller tracer region. The decrease of uncertainties of model parameters between Mock~1A and Mock~1B can be caused by local minima or specific sample of high energy stars.  It is seen from Fig.~\ref{fig:mock1_post} that selecting a smaller tracer region does not create any bias. 
	
\begin{figure}
			\includegraphics{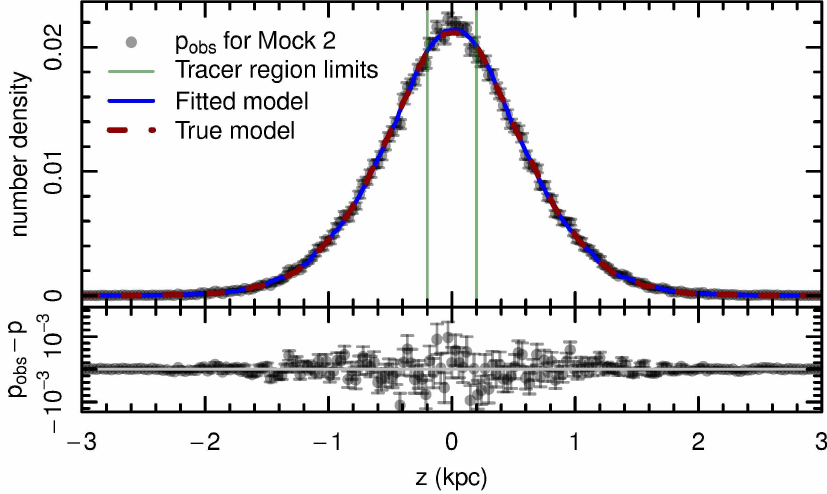}
			\caption{The fitted $p$ and $p_\mathrm{obs}$ distribution for Mock~2B. True model denotes test-particle distribution, which is calculated based on true potential parameters (see Sect.~\ref{sec:mock_simul}). The top panel shows the test-particle distribution, the bottom panel shows the residuals. }\label{fig:mock2_res}
		\end{figure}

\begin{figure}
			\includegraphics{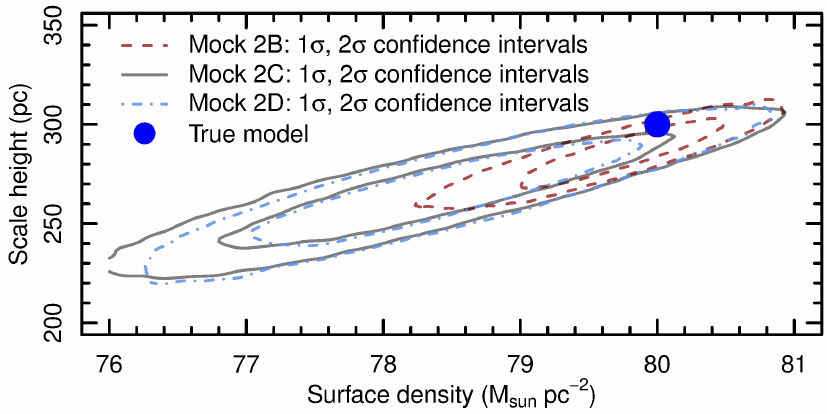}
			\caption{Posterior distributions of the parameters of the Mock~2 datasets. }\label{fig:mock2_post}
		\end{figure}

		In the third test, we compared derived results with those of Mock~2, where the test-particles were scattered much wider than in the case of Mock~1. It is seen from Figs.~\ref{fig:mock2_res} and~\ref{fig:mock2_post} that the model restores true test-particle distribution and true potential parameter values rather well. In the opposite case, where the test-particle distribution is much narrower than the changes in a given potential, the restored potential parameters have a much higher uncertainty.
	
	
		Next, as we noted in Sect.~\ref{sec:fitting}, in order to mitigate the effect of individual spikes in modelling, one can increase the errors of the $p_\mathrm{obs}$ (this is justified when working with GAIA DR1 data). This was studied by creating Mock~1C and~2C where an additional error of 25 per cent of the maximum Poisson error of the sample was added to $p_\mathrm{obs}$. In Fig.~\ref{fig:mock1_post}, we show the posterior distribution lines of Mock 1B and Mock~1C, where one can see no systematic shift, but only small increase of uncertainties. We conclude, that artificially increasing the errors of $p_\mathrm{obs}$ is a suitable way to handle the spikiness/wobbling of the model line. Subsequent mock-fittings were done with inclusion of additional $p_\mathrm{obs}$ errors. 
		
		As a last test with different Mock subsamples we studied, how much the errors of the velocities change the modelled parameters. For this we refitted Mock~2 data with randomized velocities around the true value with normal distribution and standard deviation of $2$~km\,s$^{-1}$ (Mock~2D).  The test shows, that in the case of Mock~2, the errors in velocities have negligible effect. In the case where the test-particle distribution is thinner (and the velocities are smaller), the velocity uncertainties have higher effect. For Mock~1C the velocity uncertainty created a shift in estimated surface density up to $6$~M$_\odot$\,pc$^{-2}$.
	
\begin{figure}
			\includegraphics{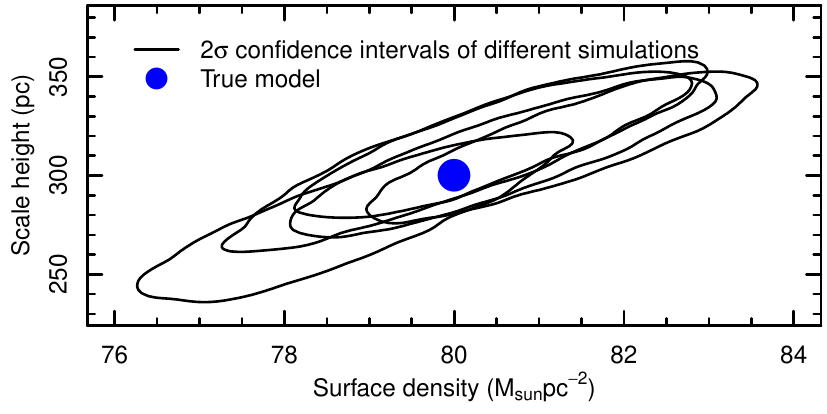}
			\caption{The parameters posterior distributions of five different simulations. The simulations have the same parameters as Mock~2C but different random seed for test particles. The figure shows variations between different runs and the effect/scatter of up-weighting low number of high-energy test-particles.}\label{fig:simul_impl}
		\end{figure}

		In addition, we tested, how much different simulation runs change the restored parameters. Fig.~\ref{fig:simul_impl} shows five different simulations, based on the same potential distribution and the same number of stars. The parameters were the same as for Mock~2C. One can see, that the true value is within two-sigma region in all cases, but the mean of the posterior distribution of values can vary somewhat. It is caused by differences of sampling of high energy stars, which were weighted up by the modelling. 

	\section{Application of the method in the Solar neighbourhood} \label{sec:application_of_solar_region}
    
By applying our method to mock data we concluded that the method allows to determine the true mass density distribution parameters with sufficient precision. In this section, we apply our method to real observational data and determine the mass density distribution in the Solar neighbourhood.

	\subsection{Data}	\label{sub:data}
    
	\subsubsection{Observations and data preparation}	\label{subsub:observations}
    
Observational data used in the present study originate from two datasets: the Tycho-Gaia Astrometric Solution (TGAS) \citep{Gaia1, Gaia2, Lindegren:2016} and the two Micron All Sky Survey (2MASS) \citep{2MASS} combined by \citet{Smart:2016}. Stellar parallaxes and proper motions were taken from TGAS, stellar $J$-band luminosities were taken from 2MASS.

	From the parallaxes ($\varpi$), the distances were calculated by adopting the the inverse value of the most likely parallax, i.e. $d = \varpi^{-1}$. 
	Thereafter, $z$ coordinates of stars will be simply as $z = d\sin\, b$, where $b$ is Galactic latitude. Velocities of stars in $z$ direction were calculated as $v_z = d\,\mu_b$, where $\mu_b$ are proper motions in Galactic latitude direction (perpendicular to the Galaxy disc). These proper motions in turn were calculated from proper motions in celestial coordinate system by using \emph{R} package \emph{astrolibR} \citep{astrolibR}. The estimated median error of the velocities $v_z$ is $3.9$~km\,s$^{-1}$. This includes the contribution of radial velocity dispersion of $\sigma_R\sin{|b|}$, where $\sigma_R$ were taken 35~$\mathrm{km}\,\mathrm{s}^{-1}$ \citep{BlandHawthorn:2016}. 
	
	To test how parallax errors affect our results, we assigned a random parallax value for each star according to the measured parallax errors while assuming that they are Gaussian. We did this 100 times and run our method for each realisation. Resulting uncertainties of modelling caused by parallax errors are described in Sect.~\ref{sub:results}.

	
	\subsubsection{Sample selection} 
	\label{subsub:selection}

	\begin{figure}
		\includegraphics[width=86mm]{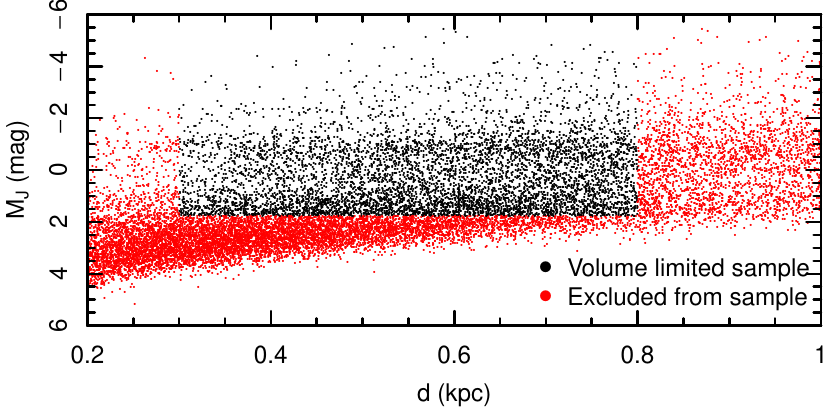}
		\caption{The subset of TGAS stars within distances of 0.2--1.0~kpc used to construct our volume limited sample. The black dots correspond to the stars forming the volume limited sample and the red dots correspond to the stars not included in the final sample.}\label{fig:sample}
	\end{figure}

	To apply our model to observed data we have to construct $p_\mathrm{obs}$ distribution and we need to define a tracer population (see Sect.~\ref{sec:tracer}). In ideal case these two datasets are identical, however, due to observational limitations tracer population should be located close to the Galaxy disc. Hence, the tracer population in current paper is just a subsample from the sample that we used for $p_\mathrm{obs}$ estimation.
 
 In our model, we assume that the density distribution depends only on the $z$ direction (distance from the Galactic plane). Hence, the stellar sample should not depend directly on distances (Malmquist bias) nor directions on the sky (dust attenuation). To tackle the dust attenuation, we construct our sample based on infrared luminosities, which are less affected by dust attenuation than optical ones. We chose 2MASS $J$-band luminosities. The apparent magnitudes were limited with $m_J < 11.0~\text{mag}$. Malmquist bias effects are controlled by constructing a volume limited sample -- only stars brighter than a certain absolute magnitude limit and closer than a certain distance were used. We limited distances by $d < 0.8~\text{kpc}$. For the given apparent magnitude limit this yields for the absolute magnitude cut-of  at $M_J = 1.8~\text{mag}$ (see Fig.~\ref{fig:sample}). To exclude possible stars that were too bright to observe with GAIA satellite, we removed all stars closer than $0.3~\mathrm{kpc}$. In addition, all sky regions, where the TGAS selection function \citep[see][]{Bovy:2017} was not defined were excluded. It mostly includes regions around ecliptic, where the TGAS is incomplete. Later the final geometry of the observed volume is taken into account while constructing the $p_\mathrm{obs}$ distribution.

Fig.~\ref{fig:sample} shows the distance and absolute luminosity cuts. Assumptions of the method were that the Galaxy is relaxed and the stellar orbits follow the overall gravitational potential, not some local minima, e.g. star clusters. Hence, we removed stars in open clusters/stellar associations. To locate stars in open clusters we used friends-of-friends (FoF) method described in \cite{Tempel:2014}. The free parameter in the cluster finding algorithm is linking length -- if two or more stars are closer than this limit, they form a cluster. We chose the linking length value as $1$~pc in the plane of sky and to suppress the distance uncertainties we chose $10$~pc for linking length in the line of sight direction. Varying the linking length values had negligible effect for the estimated $p_\mathrm{obs}$ distributions. Due to the grouping of stars, less than $3\%$ of stars were removed from our sample. The final volume limited sample contained $221\,708$ stars.

To construct $p_\mathrm{obs}$ we constructed a histogram of the $z$ positions of stars and weighted each star to take into account the effective volume of our final sample. This includes the selection effects of TGAS. Since the TGAS selection on the sky is rather complicated, we estimated the selection effects numerically. The weights were constructed as follows. First we constructed a sample of uniformly distributed points  and their $z$ distribution. Then we added the selection effects to the uniform sample by removing stars according to the spatial part of the selection function derived by \citet{Bovy:2017} and calculated their $z$ distribution. The weights were thereafter calculated by taking them to be inversely proportional to the quotient of these two $z$ distributions. We used 20 million uniformly distributed points in our survey region, hence, we were able to derive the geometrical selection function with high precision. To construct the $p_\mathrm{obs}$, we must include also the luminosity part of the selection function. It was done by combining the geometrical weight and the luminosity weight from \citet{Bovy:2017} for each star. The overall $p_\mathrm{obs}$ was constructed as weighted histogram. The errors for the $p_\mathrm{obs}$ were constructed by using bootstrap, and additional uncertainties of $0.0008$ (about $6\%$ of the mean $p_\mathrm{obs}$ value) was added to include the  systematic errors of parallaxes \citep[e.g. 0.3~mas;][]{Gaia2}. This ensures that the modelling would not fit small fluctuations of $p_\mathrm{obs}$. The resulting $p_\mathrm{obs}$ can be seen in Fig.~\ref{fig:fit_kooskola} as black points.

Next, we need to construct the tracer region. In our model, we only use the vertical velocities $v_z$. As a representative set of radial velocity data are not available to measure the exact $v_z$ for our sample of stars we restricted ourselves to proper motions data only. Due to this limit, our tracer region may include stars only from rather low Galactic latitudes $b$ as there is smallest influence to $v_z$ from other velocity components, i.e. we can assume that $v_z\simeq d\,\mu_b$. Since the modelling is sensitive to velocity errors, and the errors correlate with distances, we restricted our tracer stars by limiting the distances further to $d<0.4~\mathrm{kpc}$. This limit  effectively reduced the median of velocity errors from $3.9$ to $2.8~\mathrm{km~s}^{-1}$.
The possible centre of the Galactic plane is needed to include the lowest energy stars located at the minimum of the gravitational potential, hence the tracer population must reach the distance from the location of the Sun to the centre of Galactic plane. This distance is $25$~pc according to \citet{BlandHawthorn:2016}, hence we chose $z_\mathrm{tr}=25$~pc, which (together with the volume limited sample inner limit of $300$~pc) makes the upper limit of latitudes $|b| \le 4.8\degr$. 
To have all the velocities in the Galactic reference frame, we subtracted the Solar vertical velocity with respect to gravitational potential minima. 
We used the value of $7.2$~km\,s$^{-1}$ for the Solar motion \citep{Sperauskas:2016}. We need to weight some stars further. As the positional and luminosity part of the selection were practically independent of each other \citep{Bovy:2017}, we included the weight from under-sampling due to luminosity to the orbit weight $w_i$ in Eq.~(\ref{eq:w}). To reduce the fluctuations of very high weights, we removed the stars with luminosity weights higher than three -- it influenced total of 2\% of the sample. The size of the constructed tracer region is 4434 stars. The sample size of the tracer population cannot be increased by shifting distance or magnitude limits without affecting the precision of the velocities. 


	\subsection{The model fitting} \label{sub:fitting}
    
To derive the gravitational potential and the mass density of the Galaxy in the Solar neighbourhood we use the method described in Sect.~\ref{sec:model} and the observational data sample described in Sect.~\ref{sub:data}. We do not make any strong assumptions about the form of the overall density distribution in the Solar neighbourhood, and choose the potential as unrestricted as possible. To achieve this we approximate the density distribution with six sech$^2$ distributions in form of Eq.~(\ref{eq:sech}) which give us sufficient flexibility. Gravitational potential corresponding to one sech$^2$ density distribution is given by Eq.~(\ref{eq:sech_pot}). These components do not represent real galaxy components, but are simply a way to have an analytical but rather general and flexible expression for the gravitational potential.  The tracer was in the form of Eq.~(\ref{eq:simple_tracer}). 
		
The modelling was done by using the fitting formula (\ref{eq:ll_chisq}) by Markov Chain Monte Carlo (MCMC) with Bayesian interface. The MCMC sampling was implemented with parallel tempering to ensure the model converges to the global minimum. To suppress the uncertainties in calculations due to the edges of our volume limited sample we constrained the calculations by limiting $z$ values to $0.75~\text{kpc}$. 
The velocity uncertainties have the highest effect on  the low energy stars (i.e. the uncertainty dominates over intrinsic velocity) located in the middle of the disc with small velocity. 
As the velocities need to be precise (see Sect.~\ref{sec:model_assumptions}), we cannot use that region. If the velocity of a star located in the Galactic plane is taken to equal to the median error of the velocities ($2.8$~km\,s$^{-1}$) -- velocities that we observe for stars that are standing rest -- the star reaches heights around $|z|<70$~pc for a Milky Way like potential. Hence, to recover the overall densities we cannot model regions $|z|<70$~pc and these were excluded from the fitting process. The priors in fitting process were chosen to be the same for all components. For each component, the scale height was allowed to vary within $70~\text{pc} < h < 2000~\text{pc}$, the central location according to Gaussian distribution with zero mean and standard deviation of $250~\text{pc}$. The total central surface density was limited to be within the range $10~\text{M}_\odot \text{pc}^{-2} < \Sigma_0 < 200~\text{M}_\odot \text{pc}^{-2}$. These priors were wide enough to cover all realistic parameter values. The lower prior limit for the component heights was taken from the fitting range limit derived from the velocity error as it describes the resolution of the possible thinnest central model.

    
    \subsection{Results: calculated surface density and spatial density distributions}\label{sub:results}
		
\begin{figure}
		\includegraphics{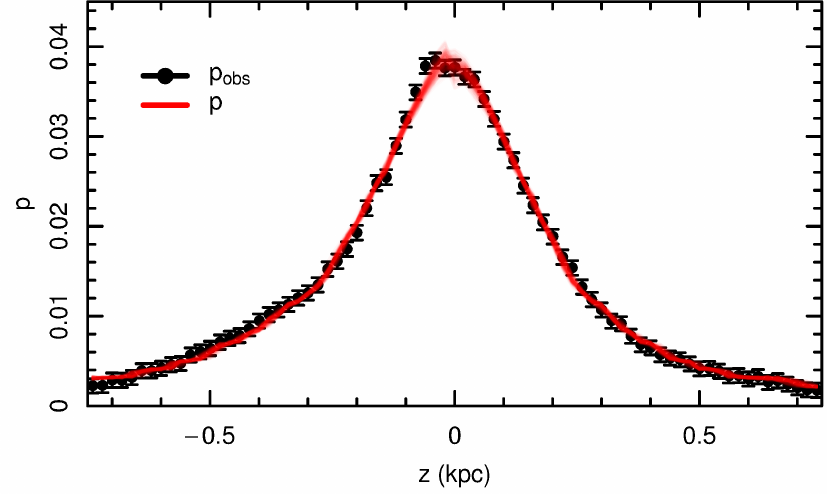}
		\caption{The observed and fitted probability density distributions with the best-fitted density-potential pair parameters. Black points with $1\sigma$ errors correspond to the density distribution of the observed sample of stars. The set of small red lines corresponds to a random sample of MCMC run points and describe the wellness of the modelling -- one can see that most of the lines are within the 1$\sigma$ range of data-points. Due to velocity uncertainties, we excluded the inner $|z|\le 70\text{pc}$ region from fitting process.} \label{fig:fit_kooskola}
\end{figure}

Calculated $p(z)$ distribution and observed $p_\text{obs}(z)$ for the best fitting model are given in Fig.~\ref{fig:fit_kooskola}. It is seen that the overall fit is rather good. 

\begin{figure}
	\includegraphics{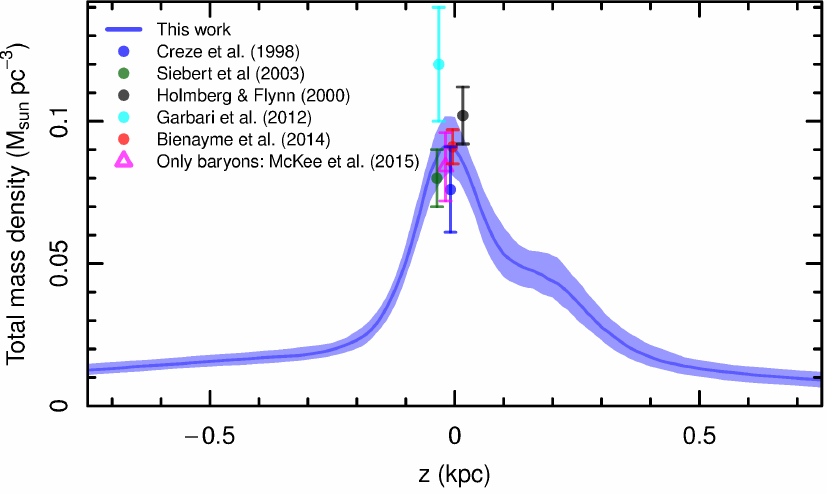}
	\caption{Calculated total matter density distribution of the model. The blue region denotes our one sigma posterior distribution of the density corresponding to the MCMC points. This contains only statistical errors of the fitting and not errors from observational uncertainties. Density values at $z = 0$ derived by other authors are given by coloured filled circles.  }\label{fig:tot_den}
	\end{figure}

One sigma posterior distribution of the calculated spatial density $\rho (z)$ taking into account observational errors is presented in Fig.~\ref{fig:tot_den} as a blue region. It is seen that the density distribution is clearly asymmetric with respect to the Galactic plane allowing to conclude that there may be some shift between different real flat components of the Galaxy.

Our calculated mass density at  $(R,z) = (R_{\odot}, 0)$ is $\rho_0 = 0.09\pm 0.01\,\text{M}_\odot \text{pc}^{-3}$. When taking into account also parallax errors then one sigma uncertainty will be $\pm 0.02\,\text{M}_\odot \text{pc}^{-3}$.

Within one sigma our result is in agreement to the values derived by \citet{Creze:1998}, \citet{Holmberg:2000}, \citet{Siebert:2003}, and \citet{Bienayme:2014} but is less than the value derived by \citet{Garbari:2012}. 
As the scatter between all these density values is within 20 per cent we may say that the old controversy from 80's between lower and higher local mass density values nearly disappeared. 

\begin{figure}
		\includegraphics{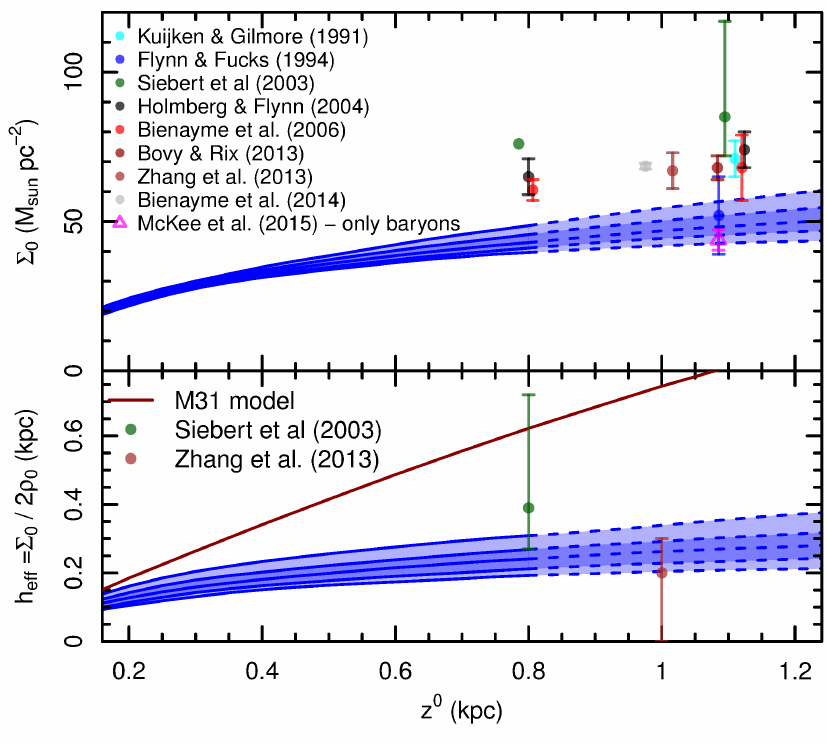}
		\caption{Surface density distribution (upper panel) and effective half-thickness (lower panel) as a function of height above the Galactic plane. The blue regions describe one and two sigma posterior distributions. This contains only statistical errors of the fitting and not errors from observational uncertainties. The dotted lines region is extrapolation of the results and thus more uncertain.} \label{fig:sigma-heff}
	\end{figure}

Another important parameter used to describe the local mass distribution is the mass surface density ($\Sigma$) at the Solar distance. For each MCMC point, we added the masses within $|z| \le z^0$ giving the total matter surface density up to the corresponding distance $\Sigma (|z| \le z^0)$. Calculated posterior distributions for $\Sigma (|z| \le z^0)$ as a function of $z^0$ up to the fitting region limit $|z| = 0.75~\text{kpc}$ are presented in Fig.~\ref{fig:sigma-heff} upper panel.  
The value for 0.75~kpc is $\Sigma (|z|\le 0.75) = 42\pm 2\,\text{M}_\odot\text{pc}^{-2}$.  When including also parallax errors then one sigma uncertainty will be $\pm 4\,\text{M}_\odot \text{pc}^{-2}$.

It is seen that our calculated surface density value is less than the values derived by \citet{Holmberg:2000}, \citet{Siebert:2003}, and \citet{Bienayme:2006} but still within two-three sigma uncertainty regions. Surface densities for $z^0 > 0.75~\text{kpc}$ were derived by extrapolating the spatial density distribution and are thus more uncertain.  But it is still seen that also our extrapolated surface density values are smaller than those calculated by other authors.

	\section{Conclusion and discussion} \label{sec:discussion}

In this paper, we presented a method to determine the matter density distribution of the Milky~Way galaxy near to the Galactic plane and tested the method with several mock data. Thereafter, the method was used with real Tycho-Gaia data.

Our method assumes that the galaxy is in a stationary state, and ignores possible gravitational potential instabilities (e.g., bending waves, warps). Non-stationarity and instabilities may cause significant systematic velocities that our method neglects. Importance of these kind of non-stationarity and instabilities requires further investigation. Vertical perturbations may bias local density determinations up to 25 per cent \citep{Banik:2017}. We did not use in the modelling process O-B stars that follow gas distribution. Thus, our results characterise overall gravitational potential distribution and are not biased due to gas distribution warps \citep{Poggio:2017}.

The method assumes that motions in $z$ and $R$ directions are decoupled; thus the third integral of motion is the $z$-directional energy. By modelling a galaxy with St\"ackel potential \citet{Statler:1989} tested validity of this assumption and concluded that this assumption can be used only up to $z \sim 1~\text{kpc}$ from the Galactic plane.  Similar result was later derived by \citet{Siebert:2008} and \citet{Smith:2009}. By using of mock N-body data \citet{Garbari:2011} derived that deviations from the assumption $I_3 = E_z$ are seen already at $z \simeq 0.5\,\text{kpc}$. 

We checked validity of this assumption in case of the Andromeda galaxy, M~31, modelled by us earlier \citep{Kipper:2016}. In case of the Andromeda galaxy at $R = 8~\text{kpc}$ the third integral can be approximated as $E_z$ up to $z\sim 0.8~\text{kpc}$ only. As this assumption is a core part to the presented method and the $I_3=E_z$  approximation may not hold away from the Galactic plane, our results should be treated with a caution.

We also calculated the local matter density given in Sect.~\ref{sub:results} when using stars only within  $|z| \le 0.5$~kpc from the Galaxy plane. We found that the results were rather similar to those found for $|z| \le 0.75$~kpc. The central matter density value $\rho_0$ remained nearly unchanged and the density distribution asymmetry was slightly reduced. 
To determine separately both the baryonic and the dark matter density distributions without using the assumption $I_3 = E_z$ more sophisticated models are needed \citep[e.g. ][]{Statler:1989, Garbari:2011, Bienayme:2014}.
 
Derived local spatial matter density value $\rho_0 = 0.09\pm 0.02~\text{M}_\odot \text{pc}^{-3}$ is only slightly larger than the estimated sum of the stellar and gas mass densities $\rho_0 = 0.084 \pm 0.012~\text{M}_\odot \text{pc}^{-3}$ \citep{Mckee:2015}. Thus, the local density of dark matter is rather small $\rho_{\text{DM}} = 0.006~\text{M}_\odot \text{pc}^{-3}$. Taking into account $2\sigma$ errors the local dark matter density would be $\le 0.038~\text{M}_\odot \text{pc}^{-3}$.

This is rather small value, meaning that when modelling the overall mass distribution of the Milky Way galaxy a maximum disc approximation can be used. Comparing this to a similar Andromeda galaxy mass distribution model \citep{Tamm:2012}, the local mass density at the distance of $R = 8~\text{kpc}$, we can notice a difference. Although the local density value in M~31 is quite similar, the best model of M~31 is not a maximum disc model and dark matter at $R = 8~\text{kpc}$ gives a quite noticeable contribution to the total density value. The same applies to the Sombrero galaxy, M~104 \citep{Tempel:2006}, but does not hold for the galaxy M~81 \citep{Tenjes:1998}. Hence, the maximum disc is not a universal approximation.

In the present paper, we computed only the total mass density, and did not distinguished between the different real Galactic components. Due to this we cannot distinguish the baryonic and dark matter density distributions. As our calculated total matter density distribution extends only up to $z = 0.75~\text{kpc}$ we cannot distinguish this from the surface density value as well. Calculated posterior distributions for the effective half-thickness $h_{\text{eff}} (|z|\le z^0) = \Sigma (|z|\le z^0) / 2 \rho (z = 0)$ as a function of $z^0$ are presented in Fig.~\ref{fig:sigma-heff} lower panel. Calculated effective half-thickness for $z^0 = 0.75~\text{kpc}$ is $h_{\text{eff}} (|z|\le 0.75) = 0.24 \pm 0.03$ kpc. This agrees  within $2\sigma$ with the values derived by \citet{Siebert:2003} and \citet{Zhang:2013}.

Derived value $h_{\text{eff}} (|z|\le 0.75) = 0.24\pm 0.03$ kpc is within the errors comparable to the observed half-thickness of baryonic stellar+gaseous matter derived by \citet{Mckee:2015} $h_{\text{eff}} (|z|\le 1.1) = \Sigma / 2 \rho = 0.26\pm 0.06$~kpc. Thus, we may conclude that all or most of the dark matter should be outside of the Galactic disc. Contribution from the dark matter with possible disc-like density distribution should be rather small or zero. This coincides with the conclusion derived from $N$-body simulations by \citet{Schaller:2016}.

It is seen from Fig.~\ref{fig:tot_den} that the density distribution is asymmetric with respect to the Galactic plane. The asymmetry is twofold. First, disc plane has a small offset by about $0.01-0.02~\text{kpc}$ and second, the densities at positive $z$ values up to $z \sim 0.4~\text{kpc}$ are systematically larger than those at corresponding negative values. 

Asymmetry in stellar density distributions between the northern and southern hemispheres of the MW have been derived earlier by e.g. \citet{Widrow:2012}, \citet{Williams:2013}, \citet{Carlin:2013}, \citet{Yanny:2013}, \citet{Xu:2015}, and \citet{Ferguson:2017}. Unfortunately, our results are not directly comparable with their results as we calculated the density distribution only at the Solar distance. Nevertheless, the order of magnitude for the disc plane offset derived by us is in agreement with the offsets 0.07~kpc and 0.14~kpc derived by \citet{Xu:2015} at distances $R = 10.5~\text{kpc}$ and 14~kpc, respectively.
 
When full Gaia observation data will be available, the method can be used at different locations and allows in this way to construct the whole 3D gravitational potential distribution in the disc. Full Gaia data release allows also to reduce errors due to incompleteness of data and thus to reduce the overall modelling uncertainties.


	\section{Acknowledgements} \label{sec:acknowledgements}
	
	We thank the referee for helpful comments and suggestions.
	We thank Indrek Kolka for useful discussion regarding the possible systematic errors in GAIA DR1 parallaxes.
	 
	This work has made use of data from the European Space Agency (ESA)
	mission {\it Gaia} (\url{http://www.cosmos.esa.int/gaia}), processed by
	the {\it Gaia} Data Processing and Analysis Consortium (DPAC,
	\url{http://www.cosmos.esa.int/web/gaia/dpac/consortium}). Funding
	for the DPAC has been provided by national institutions, in particular
	the institutions participating in the {\it Gaia} Multilateral Agreement.
    
    This publication makes use of data products from the Two Micron All Sky Survey, which is a joint project of the University of Massachusetts and the Infrared Processing and Analysis Center/California Institute of Technology, funded by the National Aeronautics and Space Administration and the National Science Foundation.
	
    This work was supported by institutional research funding IUT26-2 and IUT40-2 of the Estonian Ministry of Education and Research. We acknowledge the support by the Centre of Excellence ``Dark side of the Universe'' (TK133) financed by the European Union through the European Regional Development Fund.

	\bibliographystyle{mnras}
	\bibliography{kipper2017}

	\label{lastpage}
\end{document}